\documentclass[]{spie}  %

\usepackage{amsmath,amsfonts,amssymb}
\usepackage{graphicx}
\usepackage{algorithm}
\usepackage[noend]{algorithmic}
\usepackage{multicol}
\usepackage{wrapfig}

\usepackage[colorlinks=true, allcolors=blue]{hyperref}

\title{Secure Neuroimaging Analysis using Federated Learning with Homomorphic Encryption}

\author[a]{Dimitris Stripelis*}
\author[c]{Hamza Saleem*}
\author[c]{Tanmay Ghai*}
\author[b]{Nikhil Dhinagar*}
\author[a]{Umang Gupta}
\author[c]{Chrysovalantis Anastasiou}
\author[a]{Greg Ver Steeg}
\author[a]{Srivatsan Ravi}
\author[c]{Muhammad Naveed} 
\author[b]{Paul M.  Thompson}
\author[a]{Jos\'{e} Luis Ambite}
\affil[a]{Information Sciences Institute, University of Southern California}
\affil[b]{Imaging Genetics Center, Mark and Mary Stevens Institute for Neuroimaging and Informatics, Keck School of Medicine, University of Southern California}
\affil[c]{Viterbi School of Engineering, University of Southern California}

\authorinfo{* equal contribution. Send correspondence to D.S (stripeli@isi.edu), H.S. (hsaleem@usc.edu), P.M.T (pthomp@usc.edu) and J.L.A (ambite@isi.edu)}

\begin{document}

\maketitle              
\begin{abstract}

Federated learning (FL) enables distributed computation of machine learning models over various disparate and remote data sources, without requiring to transfer any individual sample to a centralized location. This results in improved model generalization and efficient scaling of computation as more sources and larger datasets are added to the federation. Nevertheless, recent membership inference attacks show that private or sensitive personal data can sometimes be leaked or inferred when model parameters or summary statistics are shared with a central site, requiring improved security solutions. In this work, we propose a framework for \emph{secure} FL using fully-homomorphic encryption (FHE). Specifically, we use the CKKS construction, an approximate, floating point compatible scheme that benefits from ciphertext packing and rescaling. In our evaluation on a large-scale brain MRI dataset, we use our proposed secure FL framework to train a deep learning model to predict a person's age from distributed MRI scans, a common benchmarking task, and demonstrate that there is no degradation in the learning performance between the encrypted and non-encrypted federated models. 

\end{abstract}
\section{Introduction}

Deep learning and traditional machine learning methods are now widely applied across all fields of medical imaging to facilitate image reconstruction and enhancement, automated segmentation and labeling of key structures, as well as computer aided diagnosis, pathology detection, disease subtyping, and predictive analytics (e.g., modeling future recovery or decline). In brain imaging in particular, there is great progress in automated diagnostic classification and subtyping of diseases such as Alzheimer's disease and Parkinson's disease, to assist in patient management and monitoring, and to screen patients for eligibility for clinical trials. Some recent MRI-based classifiers have merged data from over 80,000 individuals for diagnostic classification~\cite{Lu2020.08.18.256594}. 

Even so, for many tasks, publicly available training data is limited, and vast repositories of raw neuroimaging data may not be shared due to privacy concerns, as well as ethical and legal limitations on data sharing, and patient consent. In addition to local regulations that may govern or restrict data transfer, national and international laws, such as HIPAA or the General Data Protection Rule (GDPR) in Europe, severely limit the transfer of personal data across organizations and countries, even when data is anonymized as much as possible.

Federated learning offers a distributed machine learning training paradigm to perform computations on diverse datasets that may be geographically distributed across the world, or on multiple servers, by relaxing the need to centralize the data in any single location for training or testing. A federation consists of a set of learning sites (i.e., learners/clients), each performing computations on its local data, and a centralized controller that integrates information from the participating sites, and coordinates the distributed computation. The ability to learn from multiple remote data sources results in models that can generalize better than the models learned from a specific data source, while maintaining computational efficiency as the number of sites scales and datasets at each site increase in size and content.

\section{Related Work}

Several recent papers tackle the problem of learning over distributed neuroimaging datasets. The ENIGMA Consortium \cite{thompson2020enigma} has coordinated meta-analyses of statistical models from over 150 neuroimaging centers in 45 countries. Typically, parameters from models are estimated locally, and parameter estimates are shared with a central site for meta-analysis. 
Plis et al. (2016) introduced the COINSTAC system\cite{plis2016coinstac} for decentralized processing of medical imaging data, allowing iterative computations, such as N-site PCA and ICA, fitting models to local datasets and merging parameters without transferring raw data. 
Recent workshops on distributed and collaborative learning in biomedical imaging have focused on federated gradient averaging and multivariate linear modeling\cite{remedios2020federated,silva2020fed}, learning from non-IID or unbalanced data at the participating sites\cite{yeganeh2020inverse}, as well as adaptive weighting, asynchronous and semi-synchronous schemes that allow distributed learning in heterogeneous computing environments\cite{stripelis2021semi,stripelis2021scaling}. 

Other works focus on \emph{personalized} FL\cite{grimberg2020weight,chen2020fedhealth}, where models are trained on distributed data but customized to perform well on a specific dataset; in this context, the federation offers an auxiliary source of relevant data for a given task, similar in spirit to transfer learning or pre-training on auxiliary tasks. 
A similar approach based on Batch-Normalization (BN) has also been proposed\cite{andreux2020siloed}, where only the BN-transformation parameters are shared during federated training, leading to learn a per-site model instead of a single global model, and thereby accounting for heterogeneity via model personalization.
Frameworks addressing fairness in federated biomedical applications\cite{sarhan2020fairness} have also been proposed through the decomposition of locally learned representations into target and sensitive codes. FL has been applied in real-world collaborative medical settings, such as the  Breast imaging-reporting and data system (BI-RADS) tool, where a federated model for the breast density classification problem was jointly trained across seven medical institutions\cite{Roth_2020}.

\subsection{Data Privacy and Security}

Several authors have tackled data privacy and security aspects of distributed computations\cite{andreux2020siloed,gupta2021membership,plis2016coinstac}. Andreux et al \cite{andreux2020siloed} noted that even in FL, sharing aggregated statistics or model parameters can leak sensitive personal information. In a study of membership inference attacks\cite{gupta2021membership}, we recently showed that it is possible to infer if a person's data was used to train a model given only access to the model prediction (black-box) or access to the model itself (white-box), and some leaked samples from the training data distribution. We correctly identified whether specific MRI scans were used in model training with a 60\% to over 80\% success rate depending on model complexity and security assumptions. Differential privacy\cite{dwork2006differential} constitutes another proposed privacy-preserving data mining solution for biomedical settings\cite{plis2016coinstac}. However, as already shown in \cite{gupta2021membership}, even though it is an effective mechanism, it can lead to significantly increased model training costs and large drops in model performance.

\subsection{Neuroimaging Analysis}
The focus of our work is on the problem of predicting subjects' brain age from 3D MRI scans in an encrypted federated learning environment. Brain Age Gap Estimation (BrainAGE) from brain structural MRIs is a challenging biomedical task, which is an important biomarker to assess and diagnose an individual’s risk of neurological diseases. The age value difference between the
predicted and chronological brain age has been shown to be abnormally increased in a variety of neurological and psychiatric disorders, and even associated with all-cause mortality\cite{franke2019ten}. Recently, this task has been used for \emph{benchmarking}, as ground truth (the person's real age) is known. Deep learning methods have been used to predict an individual’s brain age both in centralized\cite{cole2017predicting,jonsson2019brain,lam2020accurate,peng2021accurate,gupta2021improved} and federated learning settings\cite{stripelis2021scaling}. In our study, we perform the BrainAge prediction task using a 2D Convolutional Neural Network (CNN), which was shown\cite{gupta2021improved} to yield better predictive performance compared to its 2D-Slice-RNN\cite{lam2020accurate} and 3D-CNN\cite{cole2017predicting,peng2021accurate} counterparts.

\subsection{Federated Learning}

Federated Learning was introduced in the seminal work of McMahan et al. (2017)~\cite{mcmahan2017communication}. The task was to jointly train machine learning models over mobile phones using the privately held user data without any data ever leaving the training device. The originally proposed algorithm, \textit{Federated Average} (FedAvg), assigns a fixed number of training steps to each participating device and periodically aggregates the locally trained models to compute a new community (global) model. This aggregation (synchronization) point is expressed in terms of federation rounds. Once the new community model is computed, it is distributed to all the devices and a new federation round begins. The FedAvg algorithm has been  widely adopted~\cite{smith2017federated,bonawitz2019towards,li2020federated,kairouz2019advances,yang2019federated}. In our work, we employ the FedAvg algorithm to distributively train our encrypted federated model (see section \ref{sec:FederatedLearningWithHE}). However, our architecture is general and supports other training policies.

Depending on the federated learning setting, the number of participating learners, and the amount of data, two distinct federated learning environments are commonly considered \cite{li2020federated,kairouz2019advances,rieke2020future}: the \textit{cross-device} and the \textit{cross-silo} environments. The cross-device environments characterize federations with a large number of participating learners, such as internet of things (IoT) and mobile devices, with each learner holding a relatively small amount of data, and low client availability. The cross-silo federated environments consist of a relatively small number of learners, such as organizations and data centers, with larger amounts of data and high client availability. Moreover, depending on the execution workflow and computing plan, different federated learning topologies may exist\cite{rieke2020future} with \textit{centralized} and \textit{decentralized} being the two most prominent. In the centralized case, a central entity exists (federation controller), which is responsible to coordinate the participating learners and aggregate their local models, while in the decentralized case the learners can communicate directly to each other (peer-to-peer) and aggregate models without the presence of a centralized controller. In the biomedical domain that we investigate in this work, our primary focus is on the cross-silo, centralized federated learning environments, such as a network of hospitals or a research consortium, that want to collaboratively train a machine learning model over their privately owned biomedical data.

\subsection{Homomorphic Encryption in Federated Settings}
Homomorphic Encryption (HE) has also been used in various federated settings in recent years.
Zhang et al. present BatchCrypt~\cite{batchcrypt_paper}, a system for secure FL in the cross-silo setting. They present a quantization scheme to encode weight updates as a batch of gradients, which can then be processed by leveraging single instruction multiple data (SIMD) methods. Using the Paillier scheme, they demonstrate that their technique improves the training efficiency, with a small drop in accuracy.
Truex et al.~\cite{hybrid_ppfl_paper} use both differential privacy and additive HE to train ML models in the FL setting. However, using the Paillier scheme naively for HE operations (i.e. without any optimizations, such as batching), introduces a large computational overhead.
POSEIDON~\cite{poseidon_paper} uses FHE in FL for model training. However, they consider a model in which the complete FL process, including the local training done by individual learners, is encrypted; such a formulation adds significant computational overhead. Their work suggests to combine the gradient updates using a tree-like network to reduce the learners' communication, instead of using a centralized federated learning topology where all learners interact directly with a central controller.
Ma et al.\cite{ma2021privacy} applied the CKKS scheme to FL, but 
through a multi-key HE approach, called xMK-CKKS, which employs an aggregated public key with model decryption occurring after clients share information of their individual secret keys.

\section{Federated Learning}\label{sec:FederatedLearning}

In Federated Learning settings, the learning task is to find a set of optimal parameters $w^*$ that can minimize the global objective function:
\begin{equation}\label{eq:FederatedFunction}
w^*=\underset{w}{\mathrm{argmin}} f(w) \quad\text{where}\quad f(w)=\sum_{k=1}^{N}\frac{p_k}{\mathcal{P}}F_k(w)
\end{equation}
with N representing the total number of participating learners, $p_k$ the contribution value of every learner $k$ in the federation, $\mathcal{P}=\sum p_k$ the normalization factor ($\sum_{k}^N \frac{p_k}{\mathcal{P}}=1$), and $F_k(w)$ the local objective function of learner $k$. 
We refer to the model computed using Equation~\ref{eq:FederatedFunction} as the community (global) model $w_c$.
During federated training, every learner computes its local objective function by minimizing the empirical risk over its local training dataset $D_k^T$ as $F_k(w) = \mathbb{E}_{x_k \sim D_k^T}[\ell_k(w;x_k)]$, with $\ell_k$ denoting the loss function. 
In the original FedAvg algorithm\cite{mcmahan2017communication}, the weighting (contribution) value for any learner $k$ in the federation is equal to the size of its local training set, $p_k = \left|D_k^T\right|$ and every learner minimizes its local objective function for a predefined number of steps, expressed as epochs, using Stochastic Gradient Descent (SGD) as its local solver ($\eta=$ learning rate):
\begin{equation}\label{eq:LocalSGD}
    \begin{gathered}
        w_{t+1} = w_{t} - \eta\nabla F_k(w_t)
    \end{gathered}    
\end{equation}
Once every learner completes its local training task, it shares its local model with the federation controller, and a new community model is computed.

\section{Homomorphic Encryption}

\subsection{Introduction to HE}

A homomorphic encryption (HE) scheme, unlike regular cryptographic schemes, allows for certain operations (e.g. addition, multiplication) to be performed directly over encrypted data without a need for decryption. Formally, such a scheme is \emph{homomorphic} if it satisfies the following equation: $E(m_1) * E(m_2) = E(m_1 * m_2)$ $\forall m_1, m_2 \in M$, where $*$ represents a homomorphic operation, and $M$ represents the set of all possible messages~\cite{acar2017survey}.

A HE scheme can be primarily described by 4 main algorithms: $KeyGen$, $Enc$, $Dec$, and $Eval$; they are individually discussed below:
\begin{itemize}
    \item $KeyGen(1^{\lambda}) \rightarrow (p_k, s_k)$: Takes as input the security parameter $\lambda$, and outputs a pair of keys: a public key $p_k$ and a private key $s_k$. 
    \item $Enc(p_k, m) \rightarrow c$: Takes as input the public key $p_k$, and message $m$, and outputs the ciphertext $c$.
    \item $Dec(s_k, c) \rightarrow m$: Takes as input the private key $s_k$, and ciphertext $c$ and outputs the message $m$.
    \item $Eval(p_k, F, c_{1},c_{2},..,c_{n}) \rightarrow c^{*}$: Takes as input the public key $p_k$, a permitted evaluation function $F$, and the ciphertexts $c1, ..., c_n$ and computes $F(c_{1},..,c_{n})$. The evaluation is correct if the following holds: $Dec(s_k,Eval(p_k,F,c_{1},..,c_{n})) = F(m_{1},..,m_{n})$, where $\{c_{1},..,c_{n}\}$ is the encryption of $\{m_{1},..,m_{n}\}$
\end{itemize}

The security parameter $\lambda$, informally, refers to how \emph{hard} (computationally) it is for an adversary to successfully break the encryption scheme. In general, a message $m \in M$ can be a string, integer, or other type of encoding, but for our purposes, we work with vectors of real (floating point) numbers.
Additionally, in many instances, a HE scheme can be classified as one of the following: partially-homomorphic (PHE), somewhat-homomorphic (SHE), or fully-homomorphic (FHE). PHE schemes allow for either additive or multiplicative operations over ciphertexts, SHE allow for both but up to a pre-defined limit, and FHE allow for an arbitrary number of additions and multiplications (un-bounded). In our study, we employ a fully-homomorphic encryption scheme due its inherent support for an unbounded number of additive and multiplicative operations, which are essential during the aggregation of encrypted models in federated settings.

\subsection{CKKS}\label{subsec:HECKKS}
In this work, we apply the Cheon-Kim-Kim-Song (CKKS)~\cite{ckks_paper} fully-homomorphic construction, which is based on the \emph{hardness} of the Learning-With-Error (LWE)~\cite{10.1145/1568318.1568324}, or its ring variant (RLWE~\cite{10.1145/2535925}) problem. Unlike other FHE schemes, such as the BGV~\cite{BGV_paper} and BFV~\cite{BFV_paper} (integer) constructions, CKKS allows for approximate arithmetic over real and floating point numbers. It is an approximate scheme in the sense that it provides limited (configurable) precision by treating the encryption noise as natural error incurred through approximate computations, and through dropping the least significant bits of computations via the \textit{rescaling} of encrypted data. 

Rescaling refers to the underlying process of limiting ciphertext noise, and keeping the scale (controlling the precision) constant throughout a pre-defined number of multiplications allowed in the computation. Due to this, compounded computations (specifically multiplications), scale much more efficiently compared to the integer schemes as they no longer need to be exact. Furthermore, CKKS benefits from \textit{packing}, the process of slotting multiple data values into one ciphertext, which allows for encrypted computations to be done in a Single Instruction Multiple Data (SIMD) fashion. These properties make CKKS useful for our particular use case. We refer the reader to~\cite{ckks_paper} for more specific details.

\section{Federated Learning with Homomorphic Encryption}\label{sec:FederatedLearningWithHE}

\begin{figure}[htbp]
    \centering
    \includegraphics[width=0.8\textwidth]{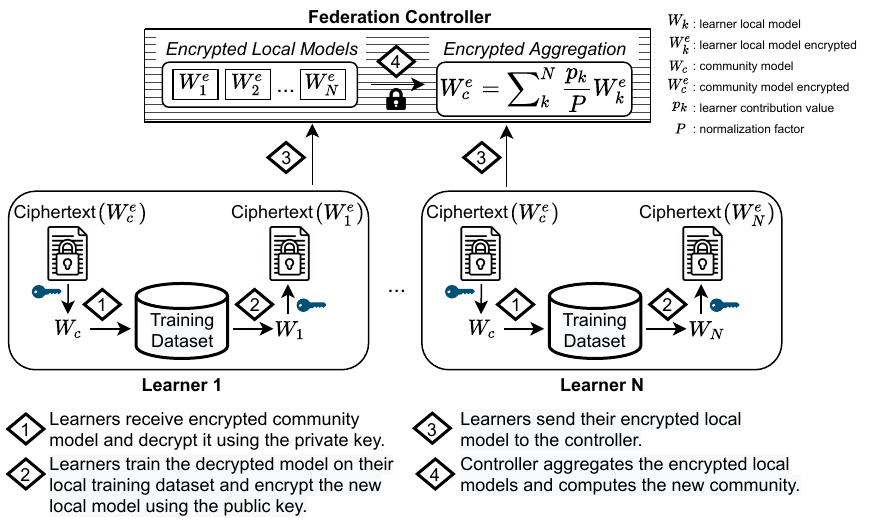}
    \caption{Federated System Architecture with Encryption}
    \label{fig:EncryptedFederatedSystemArchitecture}
\end{figure}

In a centralized federated learning environment, each learner trains on its local dataset for an assigned number of local epochs, and upon completion of its local training task, it sends the local model to the federation controller to compute the new community model. In an encrypted centralized federation environment, the procedure is similar with the addition of three pivotal key steps: \textit{encryption}, \textit{encrypted-aggregation}, \textit{decryption}. During the encryption step, every learner encrypts its locally trained model with an HE scheme using the public key, and sends the encrypted model (ciphertext) to the controller. For each learner, its encrypted model is treated as a vector of ciphertext objects, each object corresponding to a model array. With this approach, the encrypted data is represented as a (concatenated) collection of flattened data-vectors, each of them representing the local data for a particular learner. The controller receives all the encrypted local models, and then performs the encrypted weighted-aggregation to compute the new encrypted community model without ever decrypting any of the individual models. Subsequently, the controller sends the new community model to all the learners, and the learners decrypt it using the private key. Once the decryption is complete, the learners train the (new) decrypted model on their local data set and the entire procedure repeats. This pipeline is represented schematically in Fig.~\ref{fig:EncryptedFederatedSystemArchitecture}, and algorithmically in Alg.~\ref{alg:METISHE}. In our setup, the (encrypted) weighted aggregation rule applied by the controller on learners' local models is based on the local training dataset size of each learner (i.e., FedAvg). 

\begin{algorithm*}[htpb]
    \caption{\texttt{Federated Learning with HE.} Encrypted community model $w_c^e$ is computed with $N$ clients, each indexed by $k$; $\beta$ is the batch size; $\eta$ is the learning rate; $E$ are the local epochs.}
    \label{alg:METISHE}
    \begin{multicols}{2}
        \begin{algorithmic}
            \renewcommand{\algorithmicrequire}{\textbf{Initialize}: $w_c, \eta$}
            \REQUIRE
            \renewcommand{\algorithmicrequire}{\textbf{\textsc{\underline{Controller:}}}}
            \REQUIRE
            \STATE $w_c^e$ = encrypt initial community model
            \FOR{$t = 0, \dots, T-1$}
                 \FOR{each learner $k \in N$ \textbf{in parallel}}
                    \STATE send encrypted community model $w_c^e$
                    \STATE $w_k = \textsc{LearnerOpt}(w_c^e)$
                 \ENDFOR
                 \STATE $w_c^e$ = \textbf{encrypted aggregation} of all $w_k$
            \ENDFOR
            \columnbreak
            \renewcommand{\algorithmicrequire}{\textbf{\textsc{LearnerOpt($w_t^e$):}}}
            \REQUIRE        
            \STATE $w_t$ = \textbf{decrypt} community model $w_t^e$
            \STATE $\mathcal{B} \leftarrow$ Split training data $D_k^{T}$ into batches of size $\beta$
            \FOR{$i \in E$}
                \FOR{$b \in \mathcal{B}$}
                    \STATE {$w_{t+1} = w_{t} -\eta\nabla F_k(w_t;b)$}
                \ENDFOR
            \ENDFOR    
            \STATE $w_{t+1}^e =$ \textbf{encrypt} $w_{t+1}$
            \STATE send $w_{t+1}^e$ to controller
        \end{algorithmic}
    \end{multicols}
\end{algorithm*}

\section{Evaluation}
We evaluate the learning performance of FedAvg with and without encryption on the BrainAge prediction task over the UK Biobank (UKBB) neuroimaging dataset\cite{miller2016multimodal} for several data distributions environments. 

\textbf{UKBB Dataset.} For our evaluation, we use as a total of 10,446 MRI scans\cite{lam2020accurate} from subjects with no psychiatric diagnosis as defined by ICD-10 criteria out of the 16,356 available subjects in the UK Biobank dataset \cite{miller2016multimodal}. The scans were processed using a standard preprocessing pipeline with non-parametric intensity normalization for bias field correction1 and brain extraction using FreeSurfer and linear registration to a $(2 mm)^3$ UKBB minimum deformation template using FSL FLIRT, with the final dimension of the images being equal to 91×109×91. The dataset is split into a training and test sets of size 8,356, and 2,090, respectively.

\begin{figure}[htbp]
    \centering
    \includegraphics[width=1\textwidth]{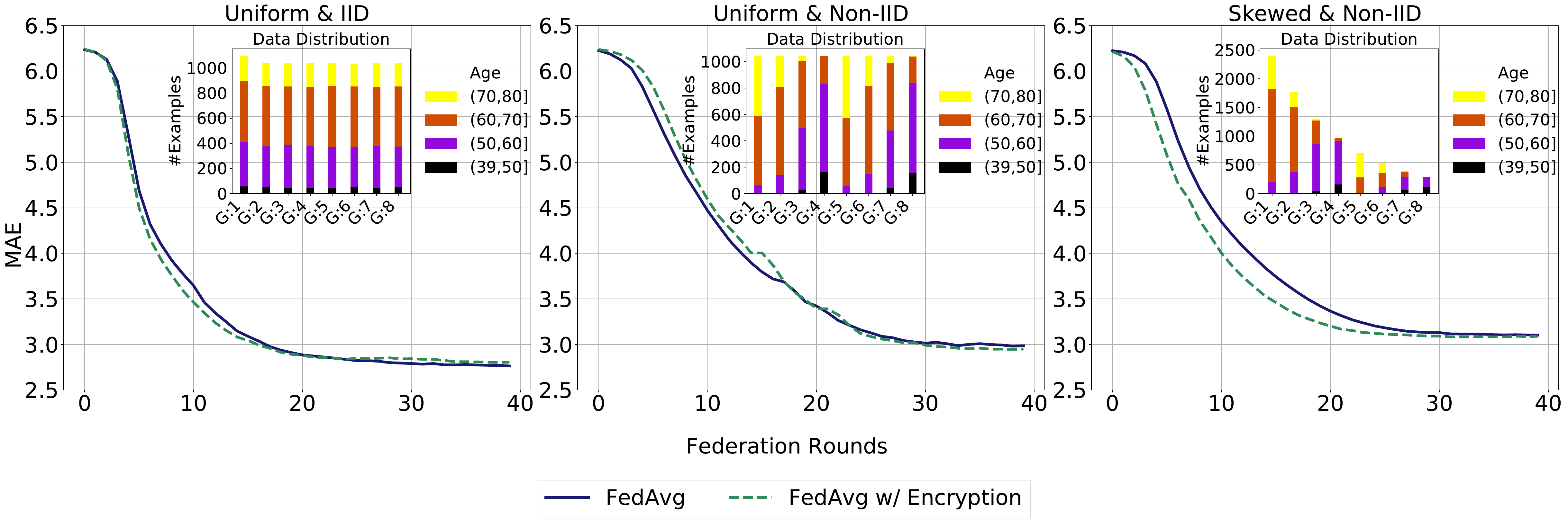}
    \caption{BrainAge Evaluation with and without Encryption}
    \label{fig:BrainAgeEvaluation}
\end{figure}

\textbf{Federation Environments.} Our evaluation testbed consists of three federated learning environments\footnote{\url{https://dataverse.harvard.edu/dataset.xhtml?persistentId=doi:10.7910/DVN/2RKAQP}} with diverse data distributions across 8 learners~\cite{stripelis2021scaling}. Each learner runs on a dedicated GPU card on a single server. 
In terms of data distribution, we investigate two distinct cases: \emph{IID}, where each learner holds training examples that follow the global data distribution, and \emph{Non-IID}, where local distributions may differ substantially from the global.  
Specifically, for the BrainAge prediction task, the IID scenario refers to the case where learners hold examples from all possible age ranges, while in Non-IID each learner holds subjects over a subset of ages. 
In terms of data amount across learners, we explore two partitioning schemes, with \emph{Uniform} representing the cases where learners hold equal number of examples, and \emph{Skewed}, the case where learners hold different amounts of examples (i.e., the assignment follows a rightly skewed distribution).
For every federated learning environment, the data distribution is provided as an inset inside the federated learning curves of Fig.~\ref{fig:BrainAgeEvaluation}. 

\textbf{CKKS Parameters.} For the implementation of the CKKS scheme, we utilize the lattice-based cryptographic \& HE library PALISADE~\cite{PALISADE}. 
In order to meet standards set by the Homomorphic Encryption Standard~\cite{cryptoeprint:2019:939}, PALISADE allows the configurability of scheme parameters that achieve certain precision, performance and security goals that a user may have. 
For our work, we configured the following parameters: \emph{multiplicative depth}=2, \emph{scale factor bits}=52, \emph{batch size}=8192, and \emph{security level}=128 bits. Multiplicative depth refers to the maximum multiplicative \emph{path}-length that may occur in a computation (e.g. multiplying $1 \cdot 2$ and then multiplying that quantity with $3$ has a path-length of 2). 
Scale factor bits represent the bit-length of the scaling factor $D$ present in the CKKS scheme; choosing it controls the bit-level accuracy of the desired computation. 
Batch size is precisely the number of plaintext slots used in a single ciphertext. 
As discussed in Section \ref{subsec:HECKKS}, CKKS can pack multiple plaintext values in each ciphertext. 
Lastly, security level controls the bit-security that such an implementation would achieve according to FHE standards. 

\begin{figure}[htbp]
    \centering
    \fbox{\includegraphics[clip, trim=1.75in 1.35in 1.95in 1.85in, width=0.5\textwidth]{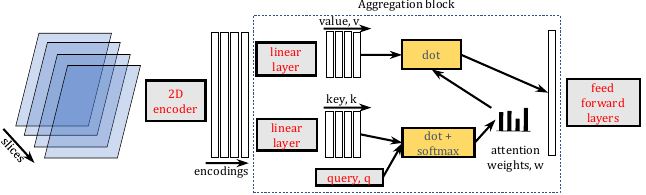}}
    \caption{2D-CNN BrainAge Model Architecture}
    \label{fig:2DCNNModelArchitecture}
\end{figure}

\textbf{Federated BrainAge Model.} Our 2D-CNN BrainAge model (1 million parameters in total) takes a 3D scan as input and encodes each slice using a 2D-CNN encoder. Subsequently, it combines the slice encodings using an attention-based aggregation module, resulting in a single embedding for the scan, which is then passed through the feed-forward layers to predict the age of the scan. An overview of the model architecture can be seen in Fig.~\ref{fig:2DCNNModelArchitecture}. During federated training, all learners receive the same model architecture and hyperparameter values, ($E=4$, $\eta = 5\times 10^{-5}$, $\beta = 1$), and use SGD as their local optimization function solver.

\textbf{Results.} We compare the convergence rate of FedAvg with and without encryption in terms of federation rounds on the three previously described federated learning environments. As shown in Fig.~\ref{fig:BrainAgeEvaluation}, encryption does not penalize the learning performance of the federated model, and in some learning scenarios (e.g., Skewed \& Non-IID) the encrypted model can lead to faster convergence compared to its non-encrypted counterpart. This can be attributed to the stochastic effect that the encryption scheme introduces to the weights of the federated learning model during encoding, decoding and private-aggregation and thereby acting as a regularizer during federated training. Eventually as training progresses, both the encrypted and non-encrypted models reach the same final score.
As the environments become more heterogeneous (from  Uniform \& IID, to  Uniform \& Non-IID, to Skewed \& Non-IID), they are harder to learn and we observe a small degradation on the final prediction error.

\vspace{-0.15cm}

\section{Discussion}

We have presented a federated learning architecture supporting fully-homomorphic encryption for privacy-preserving biomedical data analysis. 
We evaluated the performance of a non-encrypted and an encrypted federated model using the CKKS FHE scheme on the BrainAge prediction task over several data distributions of the UK BioBank dataset, showing no performance degradation under encryption. 

In our immediate future work, we plan to investigate other homomorphic encryption schemes, such as Pailler~\cite{paillier_paper}, BGV~\cite{BGV_paper}, BGN~\cite{BGN_paper}, and their respective encryption and federated training performance, as well as further adaptations that might be necessary for their successful integration into our federated training procedure (e.g., numerical scaling, protocol and data type tuning). We also plan to extend our encrypted federated neuroimaging analysis to harder learning domains, such as neurodegenerative disease prediction (e.g., Alzheimer's and Parkinson's diseases). Finally, we plan to explore how more efficient federated training protocols can be coupled with homomorphic encryption schemes to accelerate the convergence rate of the federation.

\section{Acknowledgments}
This research was supported by DARPA contract HR0011-2090104. This research has been conducted using the UK Biobank Resource under Application Number 11559.

\bibliography{main} %
\bibliographystyle{spiebib} %

\end{document}